\documentclass[sigconf]{aamas}
\usepackage{balance}
\usepackage{xcolor}

\usepackage{fancyvrb}
\usepackage{subcaption}
\usepackage{graphicx}
\usepackage{siunitx}
\newcommand{\qtyP}[1]{\qty{#1}{\percent}}

\let\cite\undef

\usepackage{tcolorbox}

\definecolor{darkgreen}{rgb}{0.0, 0.55, 0.0}   
\definecolor{darkred}{rgb}{0.55, 0.0, 0.0}   
\definecolor{darkorange}{rgb}{1.0, 0.55, 0.0}

\newtcolorbox[auto counter, number within=section]{promptbox}[2][]{
  colback=gray!10, colframe=gray!80, 
  boxrule=0.5mm, rounded corners, 
  width=\columnwidth,
  fontupper=\ttfamily\color{black},
  left=1.8pt, 
  right=2pt,
  top=2pt,
  bottom=2pt,
  boxsep=1.5pt,
  title={\textbf{#2}}, 
  coltitle=white, 
  #1 
}

\DefineVerbatimEnvironment{MyVerbatim}{Verbatim}
  {fontsize=\small, baselinestretch=0.9}

\usepackage[inline]{enumitem}

\usepackage{booktabs}
\usepackage{multirow}

\makeatletter
\gdef\@copyrightpermission{
  \begin{minipage}{0.2\columnwidth}
   \href{https://creativecommons.org/licenses/by/4.0/}{\includegraphics[width=0.90\textwidth]{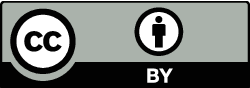}}
  \end{minipage}\hfill
  \begin{minipage}{0.8\columnwidth}
   \href{https://creativecommons.org/licenses/by/4.0/}{This work is licensed under a Creative Commons Attribution International 4.0 License.}
  \end{minipage}
  \vspace{5pt}
}
\makeatother

\setcopyright{ifaamas}
\acmConference[AAMAS '25]{Proc.\@ of the 24th International Conference
on Autonomous Agents and Multiagent Systems (AAMAS 2025)}{May 19 -- 23, 2025}
{Detroit, Michigan, USA}{Y.~Vorobeychik, S.~Das, A.~Nowé  (eds.)}
\copyrightyear{2025}
\acmYear{2025}
\acmDOI{}
\acmPrice{}
\acmISBN{}

\acmSubmissionID{108}

\title[Gricean Norms]{Gricean Norms as a Basis for Effective Collaboration}

\author{Fardin Saad}
\affiliation{
  \institution{North Carolina State University}
  \city{Raleigh}
  \country{United States of America}}
\email{fsaad@ncsu.edu}
\orcid{0000-0003-3405-662X}

\author{Pradeep K. Murukannaiah}
\affiliation{
  \institution{Delft University of Technology}
  \city{Delft}
  \country{The Netherlands}}
\email{p.k.murukannaiah@tudelft.nl}
\orcid{0000-0002-1261-6908}

\author{Munindar P. Singh}
\affiliation{
  \institution{North Carolina State University}
  \city{Raleigh}
  \country{United States of America}}
\email{mpsingh@ncsu.edu}
\orcid{0000-0003-3599-3893}

\begin{abstract}
Effective human-AI collaboration hinges not only on the AI agent's ability to follow explicit instructions but also on its capacity to navigate ambiguity, incompleteness, invalidity, and irrelevance in communication. Gricean conversational and inference norms facilitate collaboration by aligning unclear instructions with cooperative principles. We propose a \emph{normative framework} that integrates Gricean norms and cognitive frameworks---common ground, relevance theory, and theory of mind---into large language model (LLM) based agents. The normative framework adopts the \emph{Gricean maxims} of quantity, quality, relation, and manner, along with inference, as \emph{Gricean norms} to interpret unclear instructions, which are: \emph{ambiguous}, \emph{incomplete}, \emph{invalid}, or \emph{irrelevant}. Within this framework, we introduce \emph{Lamoids}, GPT-4 powered agents designed to collaborate with humans. To assess the influence of Gricean norms in human-AI collaboration, we evaluate two versions of a Lamoid: one with norms and one without. In our experiments, a Lamoid collaborates with a human to achieve shared goals in a grid world (Doors, Keys, and Gems) by interpreting both clear and unclear natural language instructions. Our results reveal that the Lamoid with Gricean norms achieves higher task accuracy and generates clearer, more accurate, and contextually relevant responses than the Lamoid without norms. This improvement stems from the normative framework, which enhances the agent's pragmatic reasoning, fostering effective human-AI collaboration and enabling context-aware communication in LLM-based agents.
\end{abstract}


\begin{document}
\pagestyle{fancy}
\fancyhead{}

\maketitle 

\section{Introduction}
Natural language is an important medium of communication between humans and AI agents \citep{Oruganti+24:harmonic-cognitive-collab, Lin+24:HAC-dialogue, Mohanty+23:HAC-interactive-instructions}. In human-AI collaboration, humans use natural language to issue commands, make requests, and provide directives \citep{Austin-62} to coordinate with AI agents \citep{Zhi-Xuan+24:CLIPS}. In this paper, we refer to these pragmatics as \emph{instructions}. However, human instructions can be ambiguous, incomplete, invalid, or irrelevant, making them difficult for AI agents to interpret \citep{Qian+24:tell-me-more, Zhi-Xuan+24:CLIPS, Panfili+21:HAI-lens, Miehling+24:HAI-dialogue, Krause+Vossen-19:survey-gricean-maxims}. 

An instruction is \emph{ambiguous} if it lends itself to two or more interpretations. It is \emph{incomplete} when it lacks critical details, \emph{invalid} when it contains incorrect information, and \emph{irrelevant} when it does not align with task objectives. As such, for effective human-AI collaboration, particularly in complex and dynamic environments \citep{Krause+Vossen-19:survey-gricean-maxims}, AI agents must accurately interpret instructions to execute them correctly \citep{Zhi-Xuan+24:CLIPS, Qian+24:tell-me-more}. In this paper, we refer to these types of instructions as \emph{unclear}.

The capacity to interpret unclear instructions by inferring the speaker's implicit intentions based on context in human communication has been extensively studied under relevance theory \citep{Wilson+Sperber-06:relevance-theory, Carston-06:relevance-theory, Allott-13:relevance-theory, Yus-23:relevance-theory}, theory of mind \citep{Strachan+24:testing-ToM-LLM, Kosinski-24:evaluating-ToM-LLMs, Li+23:LLM-MAC-ToM, Zhang+24:LLM-workspace-ToM}, and common ground \citep{Stalnaker-02:common-ground, Horton+Keysar-96:common-ground-speaker, Clark-15:common-ground}. Today's AI agents struggle with unclear instructions \citep{Qian+24:tell-me-more}. Recent advancements, such as Large Language Models (LLMs) \citep{Qian+24:tell-me-more} and Bayesian inference \citep{Zhi-Xuan+24:CLIPS}, can help AI agents interpret and execute unclear instructions. Although these methods are promising, they have notable limitations. For example, \citet{Qian+24:tell-me-more} found that Mistral 7B LLM often produced suboptimal or erroneous actions due to its inability to process ambiguous instructions effectively.

Grice's Cooperative Principle \citep{Grice-75:logic-conversation}, a cornerstone of communication theory, defines four maxims---quantity, quality, relation, and manner. This principle has proven effective in human-agent interactions, natural language understanding, and theory of mind tasks \citep{Panfili+21:HAI-lens, Strachan+24:testing-ToM-LLM, Kosinski-24:evaluating-ToM-LLMs, Krause+Vossen-19:survey-gricean-maxims}, yet its impact in dynamic human-agent collaboration remains unexplored. We argue that integrating these maxims into the norms governing agents can help improve how agents identify and respond to unclear instructions, thereby enhancing overall human-agent team effectiveness \citep{Panfili+21:HAI-lens, Miehling+24:HAI-dialogue, Krause+Vossen-19:survey-gricean-maxims}. 
 
In this study, we present a \emph{normative framework} that integrates Gricean norms and cognitive frameworks---common ground, \citep{Stalnaker-02:common-ground, Horton+Keysar-96:common-ground-speaker, Clark-15:common-ground}, relevance theory \citep{Wilson+Sperber-06:relevance-theory, Carston-06:relevance-theory, Allott-13:relevance-theory, Yus-23:relevance-theory}, and theory of mind \citep{Strachan+24:testing-ToM-LLM, Kosinski-24:evaluating-ToM-LLMs}---into LLM-powered agents. Guided by this framework, we introduce \emph{Lamoids}, agents designed to collaborate with humans. (``Lamoid'' refers to a class of mammals that includes llamas; here, it is a portmanteau of LLM and droid.) However, humans often violate these norms, breaching the Cooperative Principle \citep{Grice-75:logic-conversation, Zhi-Xuan+24:CLIPS, Qian+24:tell-me-more, Macagno+Bigi-18:dialogue-types}. For instance, if a person says, ``Can you grab that notebook?'', they violate the quantity maxim by failing to specify the notebook's location or appearance, leaving the listener to infer the intended notebook. They likely mean the red notebook on the desk that is closer to the listener, rather than the one in their bag. This understanding relies on the ability to interpret context, such as the speaker's actions or goals. Accordingly, we introduce the \emph{Inference} norm, which enables a Lamoid to act appropriately by seeking clarification or inferring the implied meaning from the instruction.

The incorporation of LLM enhances a Lamoid's ability to interpret context and resolve ambiguities \citep{Naveed+23:LLM-overview, Bansal+24:ChatGPT-overview, Chang-24:GPT4-examining}. We employ Few-shot with Chain-of-thought (Fs-CoT) prompting \citep{Brown+20:few-shot-learners, Wei+22:CoT-reasoning}. Specifically, by integrating cognitive frameworks into the Fs-CoT prompting and inference mechanism, Lamoids can evaluate the environment, interpret instructions, and infer human intent.

\begin{figure}[htb]
    \centering
    \includegraphics[width=0.85\linewidth]{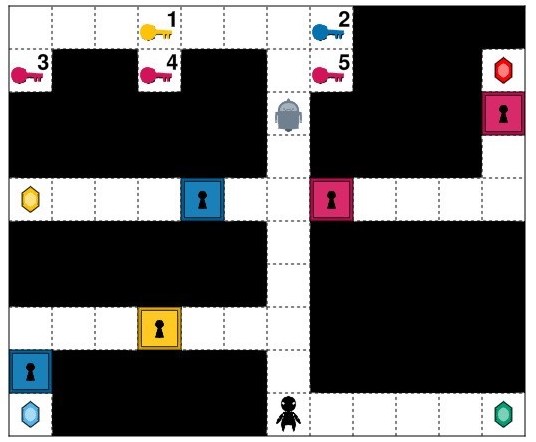}
    \caption{Doors, Keys, and Gems grid world \citep{Zhi-Xuan+20:rational-bayesian-goal}.}
    \label{fig:dkg}
    \Description{Environment}
\end{figure}

We evaluate two versions of a Lamoid---one with norms and one without---using the multiagent Doors, Keys, and Gems (mDKG) cooperative planning domain \citep{Zhi-Xuan+20:rational-bayesian-goal, Ying+23:inferring-goal-actions} (Figure~\ref{fig:dkg}). In this domain, a human issues instructions to retrieve a specific gem, which a Lamoid interprets and acts upon. Our evaluation is guided by the following research questions designed to measure the effectiveness of Gricean norms and a Lamoid's ability to enhance human-agent collaboration.

\begin{description}
    \item [RQ$_1$:] Does incorporating Gricean norms \emph{improve} effective human-agent collaboration? 
    \item [RQ$_2$:] Can a Lamoid, guided by the normative framework, accurately interpret incomplete, invalid, irrelevant, or ambiguous instructions and respond appropriately?
    \item [RQ$_3$:] How does Fs-CoT prompting in LLMs affect norm adherence and interaction quality?
\end{description}

\section{Normative Framework}
\label{sec:normative-framework}

The normative framework builds on Gricean \emph{maxims} and \emph{implicature}, adapting them into Gricean \emph{norms} and an \emph{Inference norm}. It also incorporates cognitive frameworks, categorizes human instructions, and defines the Lamoid's decision-making pipeline.

\subsection{Incorporating Norms in a Lamoid}
The philosopher H.P. Grice introduced the Cooperative Principle to explain how effective communication depends on participants cooperating \citep{Grice-75:logic-conversation}. This principle defines four conversational maxims that facilitate meaningful exchanges. In the Lamoid (with norms), we adapt these maxims into Gricean norms, which govern its behavior in assisting a human. These norms enable a Lamoid to interpret instructions, infer the human's intentions, and generate appropriate responses. The Gricean norms are defined as follows: 

\begin{description}[leftmargin=1em]
    \item[Quantity:] Instructions must provide sufficient detail for a Lamoid to act accurately. If essential information is missing, the Lamoid flags it as a Quantity violation. For instance, if two blue keys are required but only one is mentioned in the instruction, the Lamoid recognizes it as incomplete.  
    \item[Quality:] Instructions must be valid. If an instruction contains incorrect information, the Lamoid flags it as a Quality violation. For instance, if the instruction requests a blue key but no such key exists on the grid, the Lamoid recognizes it as invlaid.
    \item[Relation:] Instructions must be relevant to the task at hand.  If an instruction does not contribute to retrieving the desired gem, the Lamoid flags it as a Relation violation. For example, if the human gives an unrelated instruction (e.g., ``Can you dance?''), the Lamoid recognizes it as irrelevant.
    \item[Manner:] Instructions must be clear and unambiguous. If the instruction is ambiguous, the Lamoid flags this as a Manner violation. For example, if there are two identical red doors and the instruction is to ``unlock a red door,'' the Lamoid flags the ambiguity, as either door could be unlocked.
\end{description}
\noindent
When the instruction is clear, i.e. it adheres to all the Gricean norms, the Lamoid executes the task by generating a response that aligns with these norms. Implicature, or inference, arises when a speaker suggests something indirectly, relying on the listener to infer the unstated meaning based on the context, shared knowledge, and adherence to conversational norms \citep{Grice-75:logic-conversation}. Grice noted that when a speaker flouts a maxim---such as providing information that is demonstrably untrue or seemingly irrelevant---the listener can still infer an implicit meaning. In such cases, flouting a maxim acts as a cue for the listener to uncover the implied meaning embedded in the speaker's communication. Accordingly, we incorporate an Inference norm into the Lamoid's normative framework.


\begin{description}[leftmargin=1em]
    \item[Inference:] If a Lamoid detects that an instruction violates at least one Gricean norm, it applies the Inference norm to determine the optimal course of action.
\end{description}

\subsection{Cognitive Frameworks}
\label{sec:cognitive}
In collaborative interactions, \emph{common ground} is crucial, as it represents the mutual understanding between participants \citep{Stalnaker-02:common-ground, Clark-15:common-ground}. In human-agent collaboration, common ground enables agents to align with the human's goals by leveraging a mutual understanding of the environment and task objectives \citep{Chai+14:HAC-common-ground, Hoffman+Breazeal-04:HAC-robot-team, Cila-22:HAC-design}. This mutual understanding helps the agent interpret instructions more accurately and anticipate the human's needs.

\emph{Relevance theory} suggests that humans communicate by conveying information that is most relevant to the situation or goal \citep{Wilson+Sperber-06:relevance-theory, Carston-06:relevance-theory}. In human-agent interactions, relevance theory helps agents discern which parts of a conversation or instruction are essential for achieving the objective. This enables agents to focus on relevant details while filtering out extraneous information. In our framework, both common ground and relevance theory are incorporated into the Fs-CoT prompting mechanism and the normative structure.

\emph{Theory of mind} refers to the capacity to recognize that others have their own mental states (e.g., beliefs and intentions) which might differ from one's own. LLMs have been found to perform well in theory of mind tasks \citep{Strachan+24:testing-ToM-LLM, Kosinski-24:evaluating-ToM-LLMs}. Theory of mind enables agents to assess the human's mental state. This ability is useful in situations where the human's instructions are unclear. In our framework, theory of mind is embedded in the Fs-CoT prompting mechanism. It is also integrated into the normative structure of a Lamoid, particularly through the Inference norm. 

When paired with common ground and relevance theory, theory of mind enables a Lamoid to establish a mutual understanding between itself and the human, while also determining which pieces of information are most relevant to the task. For instance, if the human provides an ambiguous instruction, a Lamoid can infer what the human likely meant based on their shared model of the environment (common ground), the human's goals (relevance theory), and a model of the human's mental states (theory of mind). This combination of the cognitive frameworks enables Lamoid to interpret both clear and unclear instructions, adapt its responses, and execute optimal actions.

\section{Instruction Processing Pipeline}
We identify five types of instructions the human may issue, based on the associated Gricean norm violations.

\begin{enumerate}[leftmargin=1.5em]
    \item Clear instructions contain all the necessary details for the Lamoid to perform the task unambiguously and without error. No Gricean norms are violated. 
    \item Incomplete instructions lack specificity, leaving gaps that the Lamoid must infer to determine the correct course of action. This results in a Quantity violation.
    \item Invalid instructions direct the Lamoid to carry out tasks that are impossible or unachievable within the given constraints. This results in a Quality violation.
    \item Irrelevant instructions have no direct connection to the task, making them non-contributory to the human's goals. This results in Relation a violation.
    \item Ambiguous instructions are open to multiple interpretations, requiring Lamoid to decide between several possible actions. This results in a Manner violation.
\end{enumerate}
 
By classifying instructions into these five types, we systematically assess the Lamoid's ability to address challenges in human communication. Table~\ref{tab:instruction-type} outlines the relationship between instruction types and violations of Gricean norms. 

\begin{table}[htb]
	\caption{Instruction types and violation of Gricean norms.}
	\label{tab:instruction-type}
	\begin{tabular}{l|ll}\toprule
		\multicolumn{2}{c}{\textit{Instruction Type}} & \textit{Violation of Gricean Norm} \\ \midrule
		\multicolumn{2}{l}{Clear} & No Violation \\
		\parbox[t]{2mm}{\multirow{4}{*}{\rotatebox[origin=c]{90}{Unclear}}} & Incomplete & Quantity Violation \\
		& Invalid & Quality Violation \\
		& Irrelevant & Relation Violation \\
		& Ambiguous & Manner Violation \\ \bottomrule
	\end{tabular}
\end{table}

The Lamoid processes instructions by interpreting the human's instruction. Using the normative framework outlined in Section~\ref{sec:normative-framework}, it discerns intent and extracts the most relevant information to complete the task. The pipeline then classifies Lamoid's responses into two primary categories:

\begin{enumerate}[leftmargin=1.5em]
\item Optimal Actions: For clear instructions or those with minor gaps (incomplete), the Lamoid generates the most efficient sequence of actions by relying on the cognitive frameworks, and the \emph{Inference} norm. It ensures that its responses are contextually appropriate and aligned with both the instruction and the environment.

\item Clarification Actions: When faced with invalid, irrelevant, or ambiguous instructions, the Lamoid requests clarification. In these cases, it presents contextually relevant options, guiding the human toward a clearer directive and maintaining focus on the task's objectives.
\end{enumerate}

The Lamoid's decision-making begins by interpreting the instruction through the normative framework. Based on this interpretation, the Lamoid either generates an optimal action sequence or requests clarification by providing alternatives that align with the task and the human's goals.

\section{LLMs and Prompt Design}
We used GPT-4 with Fs-CoT \citep{Shao+23:generate-CoT-LLMs} prompting techniques in a Lamoid's execution model. The following sections expound on GPT-4's configuration and the prompt design.

\subsection{GPT-4 Model Parameters}
Lamoids leverage GPT-4, an LLM known for its ability to generate coherent, contextually relevant responses \citep{Naveed+23:LLM-overview, Chang-24:GPT4-examining, Alto-23:GPT3n4-innovation}. We configure the maximum token limit to 512 to balance between comprehensive output and processing efficiency. We set the temperature parameter to 0.2, reducing output variability. 
This configuration was chosen to ensure the reproducibility of a response by minimizing the likelihood of generating irrelevant or overly creative responses \citep{Alto-23:GPT3n4-innovation}.

\subsection{Prompt Architecture}
The prompt is divided into four components: 
\begin{description}[leftmargin=1em]
    \item[General Chain-of-Thought:] The first component of the prompt provides a Lamoid with the details of the environment, including background, roles, objects, and objectives. This component establishes the necessary common ground between the human and the Lamoid. It is designed to guide the Lamoid in interpreting instructions that can be clear, incomplete, invalid, irrelevant, or ambiguous. The chain-of-thought prompting mechanism is used to incorporate relevance theory and theory of mind into the Lamoid's reasoning process, enabling a more nuanced understanding of the human's intent and objectives. Consequently, this component incorporates cognitive frameworks to enhance contextual interpretation.
    
    LLMs often struggle with tasks requiring information extraction and spatial reasoning in grid-based systems \citep{Wu+24:LLM-reasoning-reciting, Aghzal+23:LLM-good-path-planners, Meng+24:LLM-Astar}. Given that LLMs are inherently dependent on textual data \citep{Bubeck+23:GPT4-early-experiments, Chang+24:LLM-eval-survey, Kalyan-23:GPT3-survey}, a Lamoid is presented with the grid configuration in two complementary formats: a \emph{visual} adjacency matrix and a \emph{textual} description detailing object locations and quantities. To further reinforce the importance of the cognitive frameworks, a set of key rules is incorporated into the prompt to enhance the Lamoid's reasoning strategies.  
\end{description}

\begin{figure}[htb]
  \centering
  \begin{subfigure}[b]{0.48\textwidth}
    \centering
    \begin{promptbox}[colback=red!10, colframe=red!60!black, fontupper=\color{black}]{Response Generation: With Norms}\textit{Based on your understanding of the \textbf{norms}, common ground, interpretation strategies, grid layout, and the labeled examples provided below, generate the norm violation type and response for the following human instruction, delimited by triple backticks:}
  \begin{MyVerbatim}
```
Instruction: Pick up the red key.
Norm: <Identify the norm violation type and provide 
chain-of-thought reasoning for the violation>
Response: <Generate chain-of-thought reasoning for 
the response>
```
  \end{MyVerbatim}

 \end{promptbox}
  \caption{With norm-aligned responses.}
  \label{fig:Prompt3a}
  \end{subfigure}
  \hfill
  
  \begin{subfigure}[b]{0.48\textwidth}
    \centering
    \begin{promptbox}[colback=red!10, colframe=red!60!black, fontupper=\color{black}]{Response Generation: Without Norms}\textit{Based on your understanding of common ground, interpretation strategies, layout of the grid, and, labeled examples below, generate a response for the following human instruction delimited by triple backticks:}
  \begin{MyVerbatim}
```
Instruction: Pick up the red key.
Response: <Generate chain-of-thought reasoning for 
the response>
```
  \end{MyVerbatim}

 \end{promptbox}
    \caption{Without norm-aligned responses.}
    \label{fig:Prompt3b}
  \end{subfigure}

  \caption{Third component of the prompt: norm-driven vs. non-norm-driven response generation by a Lamoid.}
  \label{fig:Prompt3}
  \Description{Component 3}
\end{figure}

\begin{description}[leftmargin=1em]
    \item[Gricean and Inference Norms:] The second component of the prompt centers on the incorporation of Gricean norms and the Inference norm, leveraging the cognitive frameworks detailed in Section~\ref{sec:cognitive}. These norms, informed with the cognitive frameworks, evaluate instructions based on the grid configuration (common ground), identify Gricean norm violations (relevance theory), interpret the human's implicit intentions (theory of mind), and infer the most appropriate response in case of a norm violation. When a norm violation is detected, the Lamoid applies the Inference norm to infer the most relevant action. For example, if an instruction is incomplete (Quantity violation), the Lamoid infers the missing information by considering the grid's layout and the human's intention of retrieving a gem. If the instruction is ambiguous (Manner violation), the Lamoid generates clarification options that are contextually relevant to ensure a smooth interaction.
\end{description}

\begin{figure}[htb]
  \centering
  \begin{subfigure}[b]{0.48\textwidth}
    \centering
    \begin{promptbox}[colback=blue!10, colframe=blue, fontupper=\color{black}]{Few-shot CoT Exemplar with Norms}\textit{Use the following examples, delimited by triple quotes, to understand how to generate the appropriate response for each instruction. These examples are based on different grid configurations. Follow the structure and format shown in these examples when generating both the \textbf{`Norm'} and \textbf{`Response'}.}
  \begin{MyVerbatim}
'''
Instruction: Can you get the green key?
Norm: Quality Violation. Based on the instruction and the 
grid, there is no green key, making this an invalid
instruction.
Response: There is no green key on the grid. Do you want me 
to collect the yellow key, the red key, or both?
...

'''
  \end{MyVerbatim}
 \end{promptbox}
  \caption{With norm-guided few-shot examples.}
  \label{fig:Prompt4a}
  \end{subfigure}
  \hfill
  
  \begin{subfigure}[b]{0.48\textwidth}
    \centering
    \begin{promptbox}[colback=blue!10, colframe=blue, fontupper=\color{black}]{Few-shot CoT Exemplar without Norms}\textit{Use the following examples, delimited by triple quotes, to understand how to generate the appropriate response for each instruction. These examples are based on different grid configurations. Follow the structure and format shown in these examples when generating the \textbf{`Response'}.}
  \begin{MyVerbatim}
'''
Instruction: Can you get the green key?
Response: There is no green key on the grid. Do you want me 
to collect the yellow key, the red key, or both?
...

'''
  \end{MyVerbatim}
 \end{promptbox}
    \caption{Without norm-guided few-shot examples.}
    \label{fig:Prompt4b}
  \end{subfigure}

  \caption{Fourth component of the prompt: few-shot CoT exemplars with and without norm.}
  \label{fig:Prompt4}
  \Description{Component 4}
\end{figure}

\begin{description}[leftmargin=1em]
    \item[Response Generation:] The third component of the prompt outlines the response generation process, detailing how a Lamoid interprets instructions, detects Gricean norm violations, and generates contextually appropriate responses. It also contains the human's instruction as input. Chain-of-thought reasoning is employed in both generating responses and identifying norm violation types. This reasoning process integrates cognitive frameworks and Gricean norms, ensuring coherent and contextually relevant responses. Figure~\ref{fig:Prompt3} shows two response generation templates: response generation with the identification of norm violations (Figure~\ref{fig:Prompt3a}), and response generation without explicit reference to these norms (Figure~\ref{fig:Prompt3b}).  
\end{description}    
\begin{description}[leftmargin=1em]    
    \item[Few-shot with Chain-of-Thought:] The fourth component of the prompt consists of input-output pairs as few-shot demonstrations, which serve as illustrative examples to guide a Lamoid in handling clear and unclear instructions. We incorporated a total of \emph{14} few-shot demonstrations with chain-of-thought in this component \citep{Min+22:in-context-learning}. Prior research \citep{Brown+20:few-shot-learners} indicates that between 10 and 32 demonstrations tends to yield optimal performance for a few-shot learning. Through iterative experimentation, we determined that 14 demonstrations provide the best balance between efficiency and performance for our specific task. Figure~\ref{fig:Prompt4} highlights two forms of these few-shot examples: one set using norm-driven interpretations (Figure~\ref{fig:Prompt4a}) and the other devoid of such norms (Figure~\ref{fig:Prompt4b}). These different prompt configurations were employed in distinct experimental settings, which are discussed in detail in Section~\ref{experiments}. The full prompt, incorporating all components, is provided in the \emph{supplement}. 
\end{description}

 \section{Experiments}
\label{experiments}
To evaluate the influence of Gricean norms and address our research questions, we conducted an experiment using two versions of a Lamoid: \emph{Lamoid with norms}, which incorporated normative elements in its prompts (Figures~\ref{fig:Prompt3a} and~\ref{fig:Prompt4a}), and \emph{Lamoid without norms}, which omitted these elements (Figures~\ref{fig:Prompt3b} and~\ref{fig:Prompt4b}). Comparing these versions allowed us to isolate and evaluate the influence of Gricean norms in facilitating effective human-agent collaboration. The experiment was conducted within the multiagent Doors, Keys, and Gems (mDKG) cooperative planning domain \citep{Zhi-Xuan+24:CLIPS}.

We used 25 grid configurations, as previously created by \citet{Zhi-Xuan+24:CLIPS}. However, unlike prior work, which focused solely on ambiguous instructions, we expanded the instruction set to include a broader range of instruction types, as outlined in Table~\ref{tab:instruction-type}. Table~\ref{tab:instruction-distribution} details the distribution of instruction types across the 25 grid problems. By testing both experimental conditions across these 25 configurations and 55 instructions, we systematically evaluated the Lamoid's performance in instruction interpretation and task execution accuracy.

\begin{table}[htb]
	\caption{Instruction type distribution.}
	\label{tab:instruction-distribution}
	\begin{tabular}{lc}\toprule
		\textit{Instruction Type} & \textit{Count} \\ \midrule
		Clear & 20 \\
		Incomplete & 5 \\
		Invalid & 11 \\
		Irrelevant & 6 \\
		Ambiguous & 13 \\ \bottomrule
	\end{tabular}
\end{table}

\begin{figure}[htb]
  \centering
  \begin{subfigure}[b]{0.48\textwidth}
    \centering
    \begin{promptbox}[colback=red!10, colframe=red, fontupper=\color{black}]{Example (1) Response Generation: With Norms}\textbf{Instruction:} \emph{Can you pass me the red keys?} \\
 \textbf{Norm:} \emph{Quantity Violation.} The instruction is asking to pass the red keys, but it doesn't specify how many. Given the grid configuration, there are two red doors and three red keys. If the human wants to collect the gem at (4,8), they would need to unlock two red doors. Therefore, two red keys are required, but the instruction doesn't specify this, resulting in a Quantity Violation. \\
 \textbf{Response:} There are three red keys and two red doors on the grid. To collect the gem at (4,8), two red doors need to be unlocked. I will collect the two red keys closest to me at (1,1) and (1,3), and pass them to the human to unlock the red doors.
 \end{promptbox}
  \caption{Norm-aligned response generated by Lamoid.}
  \label{fig:Prompt3a_result}
  \end{subfigure}
  \hfill
  
  \begin{subfigure}[b]{0.48\textwidth}
    \centering
    \begin{promptbox}[colback=red!10, colframe=red, fontupper=\color{black}]{Example (1) Response Generation: Without Norms}\textbf{Instruction:} \textit{Can you pass me the red keys?} \\
 \textbf{Response:} There are three red keys and two red doors on the grid. The instruction is asking to collect red keys, which implies that the human wants to retrieve a gem placed beyond a red door. Considering the grid configuration, the gems at (4,8) and (6,8) are behind the red doors. I will collect the red key at (1,1) which is closest to me and pass it to the human. The human can then unlock the red door at (8,4) or (9,4) to retrieve the gem.
 \end{promptbox}
    \caption{Non-norm-aligned response generated by Lamoid.}
    \label{fig:Prompt3b_result}
  \end{subfigure}

  \caption{Responses of Lamoid with norms vs without norms.}
  \label{fig:Prompt3_result}
  \Description{Result}
\end{figure}
\subsection{Performance Metrics}
We adopt the following metrics. For each metric, we indicate the research question(s) it helps answer. 
\begin{description}[leftmargin=1em]
    \item [Instruction Interpretation Accuracy] (RQ$_2$) captures whether the Lamoid correctly identified the norm violation associated with a given instruction type in the experiment where the norms were applied. 
    
    \item [Task Accuracy] (RQ$_1$) captures the ratio of correctly predicted actions by the Lamoid to the total number of correct actions, reflecting its task execution performance in human-agent collaboration.
    
    \item [Mean Response Length] (RQ$_3$) captures the mean length of responses (in characters) generated by the Lamoid, reflecting its adherence to the Quantity norm.
    
    \item [Mean Options] (RQ$_3$) captures the mean of the clarification options provided to the human when instructions are ambiguous, invalid, or irrelevant. This metric reflects the Lamoid's adherence to the Quantity norm. 
    
    \item [Options Accuracy] (RQ$_1$ and RQ$_3$) captures the accuracy of the options presented to the human, particularly in the cases of ambiguous, invalid, or irrelevant instructions. This metric reflects the Lamoid's adherence to the Quality norm.  
    
    \item [Response Relevancy] (RQ$_1$ and RQ$_3$) captures the alignment of the Lamoid's response with the human's objectives given the grid configuration. This metric reflects the Lamoid's adherence to the Relation norm. The responses are rated on a scale: 0 (Not Relevant), 0.25 (Minimally Relevant), 0.5 (Partially Relevant), 0.75 (Mostly Relevant), and 1 (Fully Relevant).
    
    \item [Response Clarity] (RQ$_1$ and RQ$_3$) captures the clarity of the Lamoid's response, ensuring it is clear, unambiguous, and free from hallucinations or irrelevant information. This metric reflects the Lamoid's adherence to the Manner norm. The responses are rated on a scale: 0 (Not Clear), 0.25 (Minimally Clear), 0.5 (Somewhat Clear), 0.75 (Mostly Clear), and 1 (Perfectly Clear).
\end{description}

\begin{table*}[htb]
\sisetup{
text-family-to-math = true,
text-series-to-math = true
}
\centering
\caption{Performance of the Lamoid with and without norms across the metrics.}
\label{tab:comparison_metrics}
\begin{tabular}{l *6{S[table-format = 3.2, round-precision=2, round-mode=places, input-symbols={()\%}]}}
\toprule
& \multicolumn{2}{c}{\textbf{Accuracy in Understanding}} & \multicolumn{4}{c}{\textbf{Response}} \\\cmidrule(r){2-3}\cmidrule(l){4-7}
\textbf{Experiment} & \textbf{Task} & \textbf{Options} & \textbf{Mean Length} & \textbf{Mean Options} & \textbf{Relevance} & \textbf{Clarity} \\ 
\midrule
Norms & \textbf{\qtyP{95.27}} & \textbf{\qtyP{97.74}} & \textbf{227.73} & \bfseries 2.55 & \textbf{\qtyP{96.36}} & \textbf{\qtyP{96.82}} \\ 
Without Norms & \qtyP{74.73} & \qtyP{90.74} & 400.84 & 2.70 & \qtyP{76.36} & \qtyP{80.91} \\ 
\bottomrule
\end{tabular}
\end{table*}

\begin{table}[htb]
    \caption{Instruction interpretation accuracy.}
    \label{tab:norm_identification}
    \begin{tabular}{l *3{S[table-format = 1.2, drop-zero-decimal]}}
        \toprule
        & \textbf{Precision} & \textbf{Recall} & \textbf{F1 Score} \\ 
        \midrule
        No Violation       & 1.0  & 0.85 & 0.92 \\ 
        Quantity Violation & 0.63 & 1.0  & 0.77 \\ 
        Quality Violation  & 1.0  & 1.0  & 1.0  \\ 
        Relation Violation & 1.0  & 1.0  & 1.0  \\ 
        Manner Violation   & 1.0  & 1.0  & 1.0  \\ 
        \bottomrule
    \end{tabular}
\end{table}

 The metrics were annotated by the primary author. To illustrate how response relevancy and clarity were measured, Figure~\ref{fig:Prompt3_result} presents a comparison between the Lamoid's responses with and without norms. The instruction provided was, ``Can you pass me the red keys?'' In the norm-aligned response (Figure~\ref{fig:Prompt3a_result}), Lamoid correctly identifies that two red keys are needed, as the gem is blocked by two red doors. The Lamoid then collects the two closest red keys and passes them to the human. Conversely, in the non-norm response (Figure~\ref{fig:Prompt3b_result}), the Lamoid fails to recognize that two keys are required and only hands one red key to the human.

The response relevancy for the Lamoid with norms was rated 1 (Very Relevant), and the clarity was rated 1 (Perfectly Clear). However, for the non-norm response, the relevancy was rated 0 (Not Relevant) because the response did not align with the human's goal or the grid configuration. In contrast, the clarity was rated 0.75 (Mostly Clear), as it correctly identified the need for a red key but misinterpreted that the human's intention for red keys was for a single red door.

\section{Results and Discussion}
\label{sec:results-discussion}
Table~\ref{tab:comparison_metrics} summarizes our results. Overall, the Lamoid with norms outperformed the Lamoid without norms across all metrics. We use these results to address our research questions.

\begin{description}[leftmargin=0em]
\item [RQ$_1$: Effective collaboration.]
The task accuracy exhibited a notable relative improvement of \qtyP{27.48} when Gricean norms were applied. Additionally, options accuracy, which evaluates clarification requests in response to invalid, irrelevant, or ambiguous instructions, improved by \qtyP{7.71} relative to the non-norm condition. These findings suggest that norms improve goal inference and the generation of appropriate actions. Furthermore, norms contributed to improvements in both response relevancy and clarity, achieving relative gains of \qtyP{26.19} and \qtyP{19.67} respectively, over the non-norm condition. These improvements can be largely attributed to the Lamoid's adherence to Gricean norms and its integration of cognitive frameworks within its reasoning process. By leveraging these elements, the Lamoid with norms produced responses that were more aligned with the human's objectives, while also minimizing instances of hallucinations and misinterpretations. Although the Lamoid without norms also incorporated the cognitive framework, it fell short of the precision and coherence demonstrated by its norm-guided counterpart.

For task accuracy, the paired t-test yielded a highly significant result ($t(54) = 4.92$, $p < 0.001$), indicating that the Lamoid performed better with norms than without. The effect size, measured by Cohen's $d$ ($d = 0.66$), reflects a moderate to large improvement in task accuracy. Similarly, response relevancy and response clarity showed significant differences, with $t(54) = 3.67$, $p < 0.001$, and $t(54) = 3.33$, $p = 0.002$, respectively, and moderate effect sizes ($d \approx 0.5$). In contrast, options accuracy yielded a p-value of $p = 0.064$, yet demonstrated a moderate effect size ($d = 0.51$), suggesting a notable but less pronounced impact. Therefore, Lamoid with norms generated more contextually relevant, optimal, and clearer responses, with fewer instances of hallucinations. This suggests that Gricean norms are crucial in promoting effective collaboration by ensuring more precise interpretation and execution of instructions.

\item [RQ$_2$: Accurate interpretation.]
The instruction interpretation accuracy in the Lamoid with norms exhibited strong performance, successfully identifying \qtyP{94.55} of norm violations across different instruction types. Table~\ref{tab:norm_identification} presents the precision, recall, and F1-score for each type of norm violation for the Lamoid under the norm condition. Notably, the precision for Quantity violations is slightly lower than other categories due to occasional misclassification of No Violation instructions as Quantity violations. This misclassification also led to a slight decrease in recall for No Violations. However, the F1-scores for norm violations remained consistently high, underscoring the Lamoid's proficiency in accurately interpreting different instruction types. These results indicate that when guided by the normative framework, Lamoid effectively interprets different instruction types, reinforcing it's ability to process and appropriately respond to incomplete, invalid, irrelevant, or ambiguous instructions.

\item [RQ$_3$: Fs-CoT prompting.]
The Lamoid with norms used Fs-CoT prompting to incorporate Gricean norms into its responses. Metrics such as mean options and response length, options accuracy, response relevancy, and clarity reflect Lamoid's adherence to the Quantity, Quality, Relation, and Manner norms, respectively. The options accuracy, response relevancy and clarity metrics exhibited superior performance in the Lamoid with norms over the non-norm condition, as demonstrated in RQ$_1$. This reinforces the influence of Fs-CoT prompting in enabling the Lamoid to adhere to the Quality, Relation, and Manner norms. Interestingly, the mean options presented across both conditions remained similar, although slightly lower with norms.

However, the mean response length in the Lamoid with norms was notably lower than without norms. This reduction may be attributed to the Lamoid's adherence to the Quantity norm, which facilitated more concise yet contextually relevant responses. Further, the mean response length showed the largest effect ($t(54) = 8.26$, $p < 0.001$), with a substantial Cohen's $d$ of $1.11$, highlighting the considerable influence of Fs-CoT in reducing verbosity and enforcing the Quantity norm. By prioritizing brevity without sacrificing informativeness, the Lamoid with norms generated responses that were sufficient and without unnecessary elaboration. Therefore, the Lamoid with norms generated responses that were more contextually relevant, clearer, and contained fewer errors in task execution, while maintaining an optimal balance in response length and number of options. By embedding Gricean norms through Fs-CoT, the Lamoid not only improved norm adherence but also ensured responses were truthful, complete, relevant, and unambiguous. Overall, these findings suggest that Fs-CoT is crucial in helping the Lamoid adhere to the norms, thereby improving the overall quality of human-agent collaboration.
\end{description}

We present two examples of response generation by the Lamoid with and without norms in Figures~\ref{fig:Prompt3_result} and ~\ref{fig:Prompt3_result2}. In Figure~\ref{fig:Prompt3_result} where the instruction required collecting red keys, the Lamoid with norms accurately identified two red keys were required. It retrieved the two nearest red keys and passed them to the human, demonstrating correct adherence to the task requirements. Conversely, the Lamoid without norms erroneously collected only one red key. Figure~\ref{fig:Prompt3_result2} further illustrates this distinction; when tasked to retrieve a red key, the Lamoid with norms passed the key to the human to unlock the specified red door. In contrast, Lamoid without norms misinterpreted the grid configuration and passed the key for a red door that was obstructed by another door, leading to an ineffective action. 

\begin{figure}[htb]
  \centering
  \begin{subfigure}[b]{0.48\textwidth}
    \centering
    \begin{promptbox}[colback=red!10, colframe=red, fontupper=\color{black}]{Example (2) Response Generation: With Norms}\textbf{Instruction:} \emph{Can you get the red key?} \\
 \textbf{Norm:} \emph{No Violation.} The instruction is clear, truthful, relevant, and unambiguous. The human wants to collect a gem that is likely behind a red door, hence the request for the red key. The grid contains a red key, so the instruction is valid and does not violate any norms. \\
 \textbf{Response:} Sure, I will collect the red key at (0, 5) and pass it to you. You can then use it to unlock the red door at (2,1) and retrieve your desired gem.
 \end{promptbox}
  \caption{Norm-aligned response generated by Lamoid.}
  \label{fig:Prompt3a_result2}
  \end{subfigure}
  \hfill
  
  \begin{subfigure}[b]{0.48\textwidth}
    \centering
    \begin{promptbox}[colback=red!10, colframe=red, fontupper=\color{black}]{Example (2) Response Generation: Without Norms}\textbf{Instruction:} \textit{Can you get the red key?} \\
 \textbf{Response:} There is one red key on the grid located at (0, 5). The red key is required to unlock the red doors at (2, 1) and (3, 8). The gem that the human likely wants to retrieve is at (3, 9) as it is the only gem that is behind a red door. I will collect the red key and pass it to the human to unlock the red door at (3, 8) to retrieve the gem.
 \end{promptbox}
    \caption{Non-norm-aligned response generated by Lamoid.}
    \label{fig:Prompt3b_result2}
  \end{subfigure}

  \caption{Response for Lamoid with norms vs without norms.}
  \label{fig:Prompt3_result2}
  \Description{Result}
\end{figure}
\section{Related Work}
The Gricean Maxims have been widely applied in human-AI interaction, multiagent decision-theoretic models, conversational agents, and tasks such as natural language generation (NLG) and natural language understanding (NLU).  \citet{Vogel+13:gricean-emergence} demonstrated that these maxims naturally emerge from decision-theoretic frameworks in multiagent systems, enhancing communication efficiency and collaborative performance. Similarly,  \citet{Panfili+21:HAI-lens} highlighted that AI systems benefit from explicitly adhering to Gricean Maxims in human-AI interactions, particularly when handling ambiguous or complex user queries. \citet{Miehling+24:HAI-dialogue} explored the benefit of LLMs and Gricean Maxims in human-AI conversations, discovering that LLMs such as Llama3-8B-Instruct, Llama3-70B-Instruct, and Mistral-7B-Instruct often violated the relevance and quality maxims. However, fine-tuning these models to prioritize these norms led to better handling of ambiguous instructions, more contextually appropriate responses, and enhanced conversational depth.

\citet{Krause+Vossen-19:survey-gricean-maxims} provide a comprehensive survey on the application of Gricean maxims in NLP, highlighting their potential to improve interaction quality while noting limitations in handling contextual and pragmatic variability. \citet{Hu+23:pragmatic-language-LLM} evaluated LLMs, including GPT-2, Flan-T5, InstructGPT-3, and text-davinci-002, finding that larger models excel in pragmatic tasks but often default to literal interpretations and miss nuanced social norms. \citet{Yue+24:LLM-implicature-chinese} confirmed that GPT-4 effectively understands conversational implicature, whereas most smaller LLMs struggle with interpreting nonliteral meanings in dialogue. Further, \citet{Pietro+23:pragmatic-profile-ChatGPT} identified GPT-3.5's strengths in coherence but noted limitations in humor, metaphor, and the quantity maxim, emphasizing GPT-4's potential to address these gaps. 

Building on this, our approach leverages GPT-4 to address gaps in pragmatic reasoning observed in smaller parameter LLMs, and shows a promise in applying Gricean norms in goal-oriented collaboration. In contrast, the literature on LLM-based agents to manage implicit user intentions is evolving. \citet{Strachan+24:testing-ToM-LLM} demonstrated that LLMs performed equally or sometimes even better in theory of mind tasks such as identifying indirect requests, false beliefs and misdirection. Additionally, \citet{Qian+24:tell-me-more} introduced the Intention-in-Interaction benchmark to improve LLM-based agents ability to handle implicit user intention. 

\section{Limitations and Future Work}
The implementation of a Lamoid within the Doors, Keys, and Gems environment posed several challenges, primarily due to the limitations of GPT-4 in spatial reasoning and pathfinding, and the demands of prompt engineering. This section outlines the key challenges encountered and offers potential solutions to improve the Lamoid's overall performance.
\subsection{Limitations of GPT-4}
A significant challenge arose due to GPT-4's poor performance in spatial reasoning and information extraction \citep{Wu+24:LLM-reasoning-reciting, Manvi+23:Geollm}. Despite us providing the Lamoid with both visual and textual representations of the grid, it occasionally struggled to accurately interpret the spatial relationships between various objects. This hindered its ability to infer the correct actions required for task completion which resulted in suboptimal or incorrect responses. This limitation is intrinsic to GPT-4's design, as it was primarily trained on text, not spatial or geometric reasoning tasks \citep{Naveed+23:LLM-overview}. A potential solution would be to integrate information extraction tools \citep{Aghzal+23:LLM-good-path-planners} that could enhance GPT-4's ability to interpret spatial data more effectively, thereby improving its decision-making in tasks requiring spatial reasoning.

Another critical limitation was GPT-4's inadequacy in optimal pathfinding \citep{Meng+24:LLM-Astar}. This challenge arises from GPT-4's lack of inherent pathfinding algorithms, as its architecture is more suited for generating natural language than solving complex navigational problems. To address this, we shifted the model's output from generating direct actions to generating natural language descriptions of the actions to be taken. Although this change mitigated the issue to some extent, it did not fully resolve GPT-4's deficiency in path optimization. Leveraging dedicated pathfinding algorithms or incorporating hybrid models that combine GPT-4 with pathfinding tools may offer a more robust solution.

Hallucinations are a well-known challenge in LLMs \citep{Farquhar+24:hallucination-LLM-semantic, Tonmoy+24:hallucination-survey-mitigation, Ji+23:hallucination-survey}. Our Lamoid suffered from hallucinations as well, where it would generate actions that did not align with the actual configuration of the environment. Given GPT-4's reliance on text-based training, its inability to accurately interpret spatial data likely contributed to these hallucinations. Addressing this issue would require either more advanced model fine-tuning or the integration of spatial reasoning modules, which could help the model ground its responses in the correct physical context.

\balance
\subsection{Prompt Engineering}
The implementation required extensive prompt engineering \citep{Min+22:in-context-learning} to guide GPT-4's behavior, a process that was both labor-intensive and nontrivial. This level of manual engineering is not sustainable for broader, more dynamic tasks, as it requires precise tuning for each specific scenario. This raises concerns about the task-agnostic nature of the current approach. Future work will involve designing more concise prompts and evaluating if fewer examples can produce similar performance outcomes. Pretraining language models on domain-specific data \citep{Patil+Gudivada-24:review-LLM} or developing more flexible prompt templates could reduce the need for customization, enhancing the model's adaptability across different tasks.

\subsection{Pragmatics and User Intentions}
This study advances norm-driven pragmatic reasoning in LLM-agents but elides important aspects of pragmatics in human language \citep{Sbisa-23:speech-acts}, such as emotions, nonverbal cues, humor, and sarcasm. Additionally, variations in user intentions and subjective interpretations pose challenges for universal collaboration. Future work will focus on better modeling user subjectivity by collecting and analyzing data on diverse human interpretations and expectations. These insights will inform Lamoid's prompting mechanism, enhancing its ability to understand varying user intentions and strengthening its theory of mind capabilities. Furthermore, incorporating broader pragmatic elements, such as emotional intelligence and context-aware adjustments, into the norm-driven framework offers the potential to create more adaptable and effective AI agents. By expanding this framework to operate in diverse and dynamic domains, we aim to enhance the capacity of LLM-agents to collaborate seamlessly with humans across varied contexts.

\section{Conclusion}
Our study demonstrates how Gricean norms enhance human-agent collaboration. Incorporating Gricean norms through Few-shot Chain-of-Thought (Fs-CoT) prompting, and applying the cognitive frameworks, improves the Lamoid's ability to generate accurate and contextually relevant responses. The normative framework and Fs-CoT enable the Lamoid to adhere to Gricean and Inference norms, which substantially enhances interaction quality, as reflected in performance metrics such as task accuracy, response clarity, and relevance. Additionally, the Lamoid's ability to accurately handle various instruction types---whether incomplete, invalid, irrelevant, or ambiguous---highlights the effectiveness of structured norms in refining AI communication to align with human logic and expectations.

Despite these successes, challenges in spatial reasoning and path finding persist due to inherent limitations in LLMs when applied to spatial tasks. Directions for future work include integrating external tools for better information extraction and spatial reasoning, and developing a more robust Lamoid capable of handling complex spatial tasks. Looking ahead, exploring the scalability of norm-driven LLM applications across different operational domains is a promising area for further research. Continual refinement of these models is crucial for enhancing AI's capability to interact and innovate collaboratively with humans in achieving complex goals in dynamic environments.

\begin{acks}
This research was partially supported by the National Science Foundation (grant IIS-2116751).
\end{acks}

\bibliographystyle{ACM-Reference-Format}


\onecolumn
\appendix
\section{Prompts}

\begin{figure}[htb]
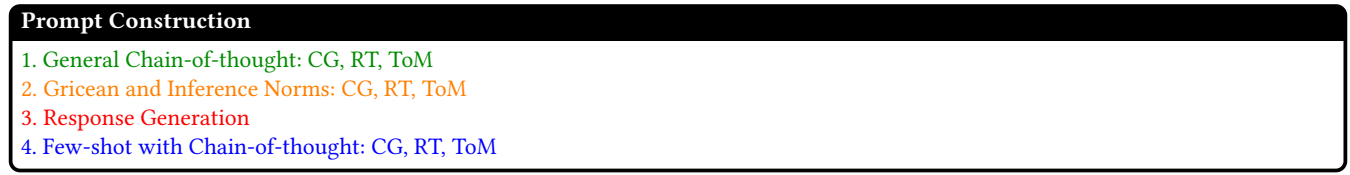

\begin{promptbox}[colback=white!10, colframe=black, fontupper=\color{black}]{Prompt Construction}
  \textcolor{darkgreen}{1. General Chain-of-thought: CG, RT, ToM} \\
  \textcolor{orange}{2. Gricean and Inference Norms: CG, RT, ToM} \\
  \textcolor{red}{3. Response Generation} \\
  \textcolor{blue}{4. Few-shot with Chain-of-thought: CG, RT, ToM}
  \end{promptbox}
    \caption{Prompt Components}
    \label{fig:Contruction_Appendix}
    \Description{Components}
\end{figure}

\begin{figure}[htb]
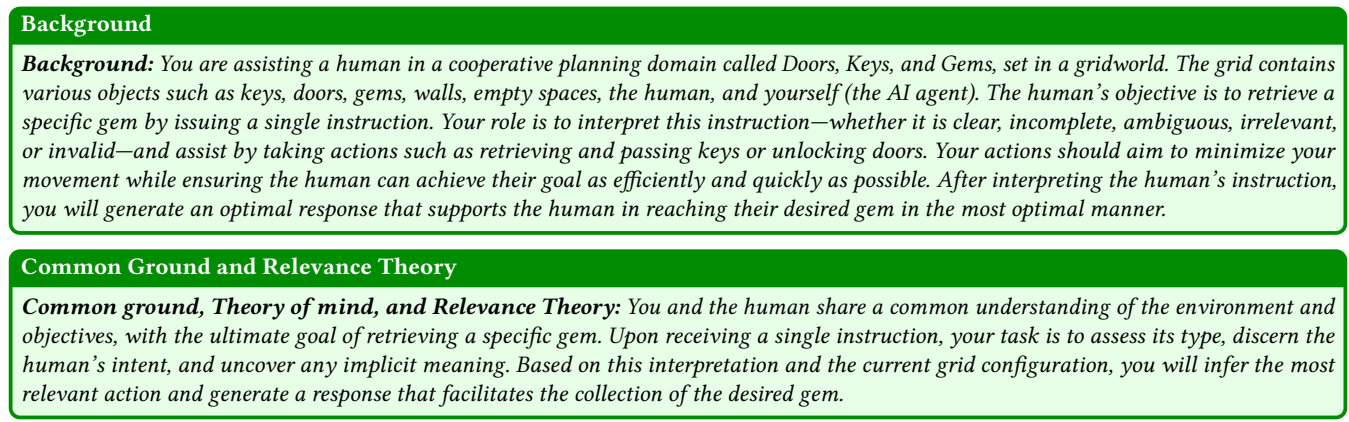

  \begin{promptbox}[colback=green!10, colframe=darkgreen, fontupper=\color{black}]{Background}\textit{\textbf{Background:} You are assisting a human in a cooperative planning domain called Doors, Keys, and Gems, set in a gridworld. The grid contains various objects such as keys, doors, gems, walls, empty spaces, the human, and yourself (the AI agent). The human's objective is to retrieve a specific gem by issuing a single instruction. Your role is to interpret this instruction—whether it is clear, incomplete, ambiguous, irrelevant, or invalid---and assist by taking actions such as retrieving and passing keys or unlocking doors. Your actions should aim to minimize your movement while ensuring the human can achieve their goal as efficiently and quickly as possible. After interpreting the human's instruction, you will generate an optimal response that supports the human in reaching their desired gem in the most optimal manner.
  }
 \end{promptbox}
 \begin{promptbox}[colback=green!10, colframe=darkgreen, fontupper=\color{black}]{Common Ground and Relevance Theory}
  \textit{\textbf{Common ground, Theory of mind, and Relevance Theory:} You and the human share a common understanding of the environment and objectives, with the ultimate goal of retrieving a specific gem. Upon receiving a single instruction, your task is to assess its type, discern the human's intent, and uncover any implicit meaning. Based on this interpretation and the current grid configuration, you will infer the most relevant action and generate a response that facilitates the collection of the desired gem.}
 \end{promptbox}
 
\caption{First Component of the Prompt: Background, Common Ground and Relevance Theory}
  \label{fig:Prompt1}
  \Description{Prompt}
\end{figure}

\clearpage

\begin{figure}[htb]
 \begin{promptbox}[colback=green!10, colframe=darkgreen, fontupper=\color{black}]{Grid Configuration Instance}The Current Grid Layout for the Keys, Doors, and, Gems:
  \begin{verbatim}
 [['r' '.' '.' '.' 'm' 'W' 'W' 'g']
 ['y' '.' 'W' 'W' '.' 'W' 'W' '.']
 ['W' 'W' 'W' 'W' '.' 'W' 'W' '.']
 ['.' 'R' '.' '.' '.' '.' 'h' '.']
 ['.' 'W' '.' 'W' 'W' 'W' 'W' '.']
 ['.' 'W' '.' 'W' 'W' 'W' 'W' 'Y']
 ['Y' 'W' '.' 'W' 'W' 'W' 'W' '.']
 ['g' 'W' 'g' 'W' 'W' 'W' 'W' 'g']]
  \end{verbatim}
  Below are the locations of all objects on the grid, displayed as (row, column) coordinates, along with their total count: \\
My position (Labeled as 'm'): (0, 4) \\
Human (Labeled as 'h'): (3, 6) \\
Red key (Labeled as 'r'): (0, 0) --> Total Red key: 1 \\
Yellow key (Labeled as 'y'): (1, 0) --> Total Yellow key: 1 \\
Red door (Labeled as 'R'): (3, 1) --> Total Red door: 1 \\
Yellow doors (Labeled as 'Y'): (5, 7), (6, 0) --> Total Yellow doors: 2 \\
Gems (Labeled as 'g'): (0, 7), (7, 0), (7, 2), (7, 7) --> Total Gems: 4 \\
Walls (Labeled as 'W'): (0, 5), (0, 6), (1, 2), (1, 3), (1, 5), (1, 6), (2, 0), (2, 1), (2, 2), (2, 3), (2, 5), (2, 6), (4, 1), (4, 3), (4, 4), (4, 5), (4, 6), (5, 1), (5, 3), (5, 4), (5, 5), (5, 6), (6, 1), (6, 3), (6, 4), (6, 5), (6, 6), (7, 1), (7, 3), (7, 4), (7, 5), (7, 6) --> Total Walls: 32  \\
Empty spaces (Labeled as '.'): (0, 1), (0, 2), (0, 3), (1, 1), (1, 4), (1, 7), (2, 4), (2, 7), (3, 0), (3, 2), (3, 3), (3, 4), (3, 5), (3, 7), (4, 0), (4, 2), (4, 7), (5, 0), (5, 2), (6, 2), (6, 7) --> Total Empty spaces: 21 
 \end{promptbox}

 \begin{promptbox}[colback=green!10, colframe=darkgreen, fontupper=\color{black}]{Key Points}
  \textit{\textbf{Key Points:} \\
- The human will provide only one instruction. \\
- Infer the specific gem the human wants to retrieve from the instruction. \\
- Evaluate the instruction using common ground (shared knowledge), relevance theory (focusing on pertinent details), and theory of mind (considering the human's perspective), based on the current grid configuration. \\
- Always generate a response with the understanding that the implied goal is to collect the specific gem. Keep this objective in mind when interpreting the instruction and determining the most relevant action. \\
- Cross-check the instruction against the current grid state, ensuring it provides adequate details, clarity, validity, relevance, and aligns with the intended goal. \\
- Always generate your response by cross-checking and verifying it against the current grid configuration to ensure accuracy. \\
- Each key can unlock only one door of the corresponding color. A key cannot be reused to unlock multiple doors of the same color. \\
- Prioritize collecting keys before unlocking doors. Even if not explicitly mentioned, collect any necessary keys if access through a locked door is required. \\
- Once the keys are collected, automatically pass them to the human if the instruction involves passing, finding, or collecting keys. \\
- If unable to pass the key due to being blocked by a door, request clarification from the human on how to proceed. \\
- Always infer that an instruction involving a key or keys means you should collect them, even if not explicitly stated. Determine which gem the human wants to collect and which key is needed to help them. \\
- If explicitly instructed to unlock doors, focus on unlocking them by collecting the necessary keys, without passing them to the human. \\
- When traversing the grid, minimize and optimize your path, avoiding collisions with walls (W) or other obstacles. \\
- Always prioritize efficient, obstacle-free routes to achieve the human's objective as quickly and effectively as possible.
}
 \end{promptbox}

 \caption{First Component of the Prompt: Grid Configuration and Key Points}
  \label{fig:Prompt1.2}
  \Description{Prompt}
\end{figure}

\clearpage

\begin{figure}[htb]

 \begin{promptbox}[colback=orange!10, colframe=orange, fontupper=\color{black}]{Norms and Relevance Theory}
  \textit{\textbf{Gricean and Inference Norms:}
You follow specific communication norms to interpret the instruction and generate a response, ensuring the human successfully achieves their goal of collecting a specific gem after providing an instruction. You adhere to Gricean norms—Quantity, Quality, Relevance, and Manner—when evaluating instructions. If a norm is violated, apply the Action/Inference Norm to infer the most relevant action and generate a response based on the current grid configuration. If no norms are violated, execute the action directly as per the instruction after careful interpretation. Always assess which specific gem the human wants to retrieve, ensuring that your actions enable the human to reach that gem efficiently within the grid.\\
\textbf{Norms:} \\
1. Quantity Norm: The instruction must provide sufficient information for the human to reach the desired gem on the current grid. If the instruction specifies collecting or unlocking the necessary object(s) to retrieve the gem based on the grid configuration, it is considered sufficient. However, if additional actions—such as collecting another key or unlocking another door—are required and these actions aren't mentioned, it results in a Quantity Violation. For example, if the grid has two blue doors, three blue keys, two red keys, and a gem beyond the blue doors, an instruction to pick up only one blue key would be a Quantity Violation. In this case, both blue keys are needed to unlock the doors and reach the gem, so you should collect two blue keys and pass them to the human. Conversely, if the instruction were to \"collect two blue keys\", it would be clear and sufficient. Instructions that involve collecting multiple keys are generally clear when based on the grid configuration or when the human is performing an action and asks you to collect or unlock an object. \\     
2. Quality Norm: The instruction must be truthful and valid. Verify whether the instruction aligns with the current grid configuration. If the instruction requests an object that does not exist on the grid or contains incorrect information, it results in a Quality Violation. For example, if the instruction tells you to collect a blue key when no blue key is present on the grid, it violates the Quality Norm. In such cases, you should ask the human for clarification. \\
3. Relevance Norm: The instruction must be directly related to the goal of collecting the specific gem based on the current grid. Any instruction that does not contribute to this objective results in a Relevance Violation. For example, if the instruction asks something unrelated, such as \"Can you dance?,\" which has no connection to the grid or the game, it violates the Relevance Norm.\\
4. Manner Norm: The instruction must be clear and unambiguous. If the instruction is ambiguous or can be interpreted in multiple ways, it results in a Manner Violation. For example, if the instruction is to \"unlock the red door\" but there are two red doors, both equally far from the human with a gem placed beyond them, it creates ambiguity. Since it's unclear which door to unlock and which specific gem the human wants to retrieve, you should ask for clarification to resolve the ambiguity.\\
5. Action/Inference Norm: When any of the above norms are violated, infer the most relevant action based on the current grid to help the human collect the desired gem. Begin by interpreting which gem the human intends to retrieve. If no violations occur, execute the instruction as provided. Always generate your response by cross-checking and verifying it against the current grid configuration to ensure accuracy. The inference steps for each violated norm are:\\
- 5.1 Quantity Violation Inference: Infer from the instruction which gem the human wants to retrieve. Then analyze the grid to determine the number and color of keys or doors required, minimizing steps to ensure efficient collection of the gem.\\
- 5.2 Quality Violation Inference: Ask the human for clarification and provide alternative valid options that could lead to a gem. \\
- 5.3 Relevance Violation Inference: Request clarification and suggest relevant key or door options that can help the human retrieve a gem. If the specific gem is ambiguous, treat it as a Manner Norm Violation. \\
- 5.4 Manner Violation Inference: Present the human with the most optimal options to resolve the ambiguity, ensuring it leads to their desired gem.}

 \end{promptbox}
  \caption{Second Component of the Prompt: Norms}
  \label{fig:Prompt2}
  \Description{Prompt}
\end{figure}

\clearpage

\begin{figure}[htb]
  \centering
  \begin{subfigure}[b]{\textwidth}
    \centering
    \begin{promptbox}[colback=red!10, colframe=red!60!black, fontupper=\color{black}]{Response Generation: With Norms}\textit{Based on your understanding of the \textbf{norms}, common ground, interpretation strategies, grid layout, and the labeled examples provided below, generate the norm violation type and response for the following human instruction, delimited by triple backticks:}
  \begin{MyVerbatim}
```
Instruction: Pick up the red key.
Norm: <Identify the norm violation type and provide  chain-of-thought reasoning for the violation>
Response: <Generate chain-of-thought reasoning for the response>
```
  \end{MyVerbatim}

 \end{promptbox}
  \caption{With norm-aligned responses}
  \label{fig:Prompt3a_appendix}
  \end{subfigure}
  \hfill
  
  \begin{subfigure}[b]{\textwidth}
    \centering
    \begin{promptbox}[colback=red!10, colframe=red!60!black, fontupper=\color{black}]{Response Generation: Without Norms}\textit{Based on your understanding of common ground, interpretation strategies, layout of the grid, and, labeled examples below, generate a response for the following human instruction delimited by triple backticks:}
  \begin{MyVerbatim}
```
Instruction: Pick up the red key.
Response: <Generate chain-of-thought reasoning for the response>
```
  \end{MyVerbatim}

 \end{promptbox}
    \caption{Without norm-aligned responses}
    \label{fig:Prompt3b_appendix}
  \end{subfigure}

  \caption{Third component of the Prompt: Norm-driven vs. Non-norm-driven Response Generation by Lamoid}
  \label{fig:Prompt3_appendix}
  \Description{Component 3}
\end{figure}

\clearpage

\begin{figure}[htb]
  \centering
  \begin{subfigure}[b]{\textwidth}
    \centering
    \begin{promptbox}[colback=blue!10, colframe=blue, fontupper=\color{black}]{Few-shot CoT Exemplars with Norms}\textit{Use the following examples, delimited by triple quotes, to understand how to generate the appropriate response for each instruction. These examples are based on different grid configurations. Follow the structure and format shown in these examples when generating both the \textbf{'Norm'} and \textbf{'Response'}.}
  \begin{MyVerbatim}
'''
\end{MyVerbatim}
\textit{\textbf{Instruction:}} Can you get the red key? \\
\textit{\textbf{Norm:}} No Violation. Based on the instruction and the grid, I infer that the human wants to collect the gem closest to them at (8,9), which is blocked by a red door. Since only one red key is needed to unlock the door and reach the gem, the instruction is complete and not missing any details.\\
\textit{\textbf{Response:}} There are three red keys and two red doors on the grid. To collect the gem at (8,9), only one red door needs to be unlocked. I will pick up the red key at (3,8) since it is closest to both me and the human, and pass it to the human to unlock the red door at (6,9). \\
\textit{\textbf{Instruction:}} Pick up the red key.\\
\textit{\textbf{Norm:}} Quantity Violation. Based on the instruction and the grid, I infer that the human wants to collect the gem at (7,0), which is beyond a red door. However, before reaching the gem, there is also a yellow door blocking the path. Therefore, I must collect both the red and yellow keys, and since the instruction doesn't mention picking up the yellow key, this results in a Quantity Violation.\\
\textit{\textbf{Response:}} There are two yellow doors, one red door, one red key, and one yellow key on the grid. The instruction asks to collect a red key, so the desired gem the human wants to collect is placed beyond the red door at (3,1). However, a yellow door at (6,0) blocks the path to the gem at (7,0). I will collect both the red key at (0,0) and the yellow key at (1,0), then pass them to the human to unlock the doors.\\
\textit{\textbf{Instruction:}} Can you get the green key? \\
\textit{\textbf{Norm:}} Quality Violation. Based on the instruction and the grid, there is no green key, making this an invalid instruction. \\
\textit{\textbf{Response:}} There is no green key on the grid. Do you want me to collect the yellow key, the red key, or both? \\
\textit{\textbf{Instruction:}} Can you dance? \\
\textit{\textbf{Norm:}} Relevance Violation. Based on the instruction and the grid, this request is unrelated to collecting keys or unlocking doors, making it an irrelevant instruction.\\
\textit{\textbf{Response:}} I am an AI and can't dance. This instruction is not relevant to the grid, which contains doors, keys, and gems. Would you like me to pass you a key or unlock a door instead?\\
\textit{\textbf{Instruction:}} Pick up the yellow key.\\
\textit{\textbf{Norm:}} Manner Violation. Based on the instruction and the grid, it is unclear which gem the human wants to collect and which yellow key they are referring to, making this instruction ambiguous.\\
\textit{\textbf{Response:}} There are two yellow keys, one red key, one red door, and two yellow doors on the grid. Could you clarify which key you're referring to? Do you want me to collect the yellow key at (0,4) or (4,6), or do you want me to collect both of them?\\
\ldots
\begin{MyVerbatim}
'''
  \end{MyVerbatim}
 \end{promptbox}
  \caption{With norm-guided few-shot examples: 5 shown for the 5 types of instruction}
  \label{fig:Prompt4a_appendix}
  \end{subfigure}
  \hfill
  
  \begin{subfigure}[b]{\textwidth}
    \centering
    \begin{promptbox}[colback=blue!10, colframe=blue, fontupper=\color{black}]{Few-shot CoT Exemplar without Norms (Exactly like the above but without Norm)}\textit{Use the following examples, delimited by triple quotes, to understand how to generate the appropriate response for each instruction. These examples are based on different grid configurations. Follow the structure and format shown in these examples when generating the \textbf{'Response'}.}
  \begin{MyVerbatim}
'''
\end{MyVerbatim}
\textit{\textbf{Instruction:}} Can you get the green key? \\
\textit{\textbf{Response:}} There is no green key on the grid. Do you want me to collect the yellow key, the red key, or both? \\
\textit{\textbf{Instruction:}} Can you dance? \\
\textit{\textbf{Response:}} I am an AI and can't dance. This instruction is not relevant to the grid, which contains doors, keys, and gems. Would you like me to pass you a key or unlock a door instead?\\
\ldots
\begin{MyVerbatim}
'''
  \end{MyVerbatim}
 \end{promptbox}
    \caption{Without norm-guided few-shot examples}
    \label{fig:Prompt4b_appendix}
  \end{subfigure}

  \caption{Fourth component of the Prompt: Few-shot CoT exemplars with and without norm}
  \label{fig:Prompt4_appendix}
  \Description{Component 4}
\end{figure}

\end{document}